\documentstyle[12pt,amssymb,amsfonts,amsbsy]{article}
\begin{document}
\author{Ilja Schmelzer\thanks
       {WIAS Berlin}}

 \title{A Metric Theory of Gravity with Condensed Matter
 Interpretation}
\sloppypar 

\begin{abstract}
We present a metric theory of gravity with Lagrangian

\[L =  (8\pi G)^{-1}(\Xi g^{ii} - \Upsilon g^{00})\sqrt{-g}
    +  L_{GR} + L_{matter}\]

motivated by classical equations

\begin{eqnarray*}
\partial_t \rho + \partial_i (\rho v^i) &= &0 \\
\partial_t (\rho v^j) + \partial_i(\rho v^i v^j + p^{ij}) &= &0
\end{eqnarray*}

for a medium in Newtonian space-time.  We obtain stable ``frozen
stars'' instead of black holes and a ``big bounce'' instead of a big
bang singularity.

\end{abstract}

\maketitle

\section{Introduction}

There is a close analogy between condensed matter theory and gravity.
It has been recognized that ``effective gravity, as a low-frequency
phenomenon, arises in many condensed matter systems'' \cite{Volovik}.
This has been used to study Hawking radiation and the Unruh effect
\cite{Visser} \cite{Unruh} \cite{Jacobson} \cite{Volovik} and vacuum
energy \cite{Volovik} for condensed matter examples.  The general
exchange of ideas with high energy physics, which ``includes global
and local spontaneous symmetry breaking, the renormalization group,
effective field theory, solitons, instantons, and fractional charge
and statistics'' \cite{Wilczek}, is also worth to be mentioned.

The theory of gravity we present here suggests that this is not an
accident, but the gravitational field is a medium in Newtonian
space-time, described by usual condensed matter variables, with an
interesting Lagrange formalism.  Few general assumptions are
sufficient to obtain a Lagrangian very close to GR, which fulfills the
Einstein equivalence principle:

\[L =  (8\pi G)^{-1}(\Xi g^{ii} - \Upsilon g^{00})\sqrt{-g}
    +  L_{GR} + L_{matter}\]

After the derivation of the theory we consider quantization, some
remarkable predictions, and compare the theory with other theories of
gravity.

\section{The Theory}

The theory describes a classical medium in a Newtonian framework --
Euclidean space and absolute time.  But we prefer to present the
theory in a formalism where the non-covariant terms are disguised as
covariant, with the preferred coordinates considered as usual scalar
fields $X^\mu(x)$.  It is easy to transform a non-covariant Lagrangian
$L=L(T^{\ldots}_{\ldots},\partial_\mu T^{\ldots}_{\ldots})$ into a
(formally) covariant form
$L=L(T^{\ldots}_{\ldots},\partial_{\mu}T^{\ldots}_{\ldots},X^\mu_{,\nu})$.

The medium is described by steps of freedom typical for condensed
matter theory.  The gravitational field is defined by a positive
density $\rho$, a velocity $v^i$, and a negative-definite symmetrical
tensor field $p^{ij}$ which we name ``pressure''.  The effective
metric $g_{\mu\nu}$ is defined algebraically by

\begin{eqnarray*}\label{gdef}
 \hat{g}^{00} = g^{00} \sqrt{-g} &=  &\rho \\
 \hat{g}^{i0} = g^{i0} \sqrt{-g} &=  &\rho v^i \\
 \hat{g}^{ij} = g^{ij} \sqrt{-g} &=  &\rho v^i v^j + p^{ij}
\end{eqnarray*}

This decomposition of $g^{\mu\nu}$ into $\rho$, $v^i$ and $p^{ij}$ is
a variant of the ADM decomposition.  The signature of $g^{\mu\nu}$
follows from $\rho>0$ and negative definiteness of $p^{ij}$.

The theory does not specify all properties of the medium, but only a
few general properties -- especially the conservation laws.  The
``material properties'' of the medium, denoted by $\varphi^m$, remain
unspecified.  They are the matter fields.  The complete specification
-- which includes the material laws of the medium -- gives the theory
of everything.  The few general properties fixed here define a theory
of gravity similar to GR.  While it leaves the matter steps of freedom
and the matter Lagrangian unspecified, it derives the Einstein
equivalence principle.

In our covariant formalism the conservation laws may be defined as the
Euler-Lagrange equations for the preferred coordinates.  The related
energy-momentum tensor

\[ T^\nu_\mu = - {\partial L\over\partial X^\mu_{,\nu}} \]

is not the same as in Noether's theorem, but only equivalent.  Now, we
identify these conservation laws with the conservation laws we know
from condensed matter theory.  First, the Euler-Lagrange equation for
time we identify with the classical continuity equation for the
medium:

\begin{equation} \label{continuity}
 \partial_t \rho + \partial_i (\rho v^i) = 0
\end{equation}

The equations for the spatial coordinates we identify with the Euler
equation:

\begin{equation} \label{momentum}
 \partial_t (\rho v^j) + \partial_i(\rho v^i v^j+p^{ij})=0
\end{equation}

Note that we use here the identification of matter fields with
material properties of the medium -- we have no momentum exchange with
external matter.  The four conservation laws transform into the
harmonic condition for the metric $g_{\mu\nu}$.  Thus, they really
look like equations for the preferred coordinates:

\[ \Box X^\nu = \partial_\mu (g^{\mu\nu}\sqrt{-g}) = 0\]

Therefore, we assume that the conservation laws are proportional to
the Euler-Lagrange equations of $S = \int L$ for the preferred
coordinates $X^\mu$:

\[{\delta S\over\delta X^\mu}\equiv-(4\pi G)^{-1}\gamma_{\mu\nu}\Box X^\nu\]

We have introduced here a constant diagonal matrix $\gamma_{\mu\nu}$
and a common factor $-(4\pi G)^{-1}$ to obtain appropriate units.
With Euclidean symmetry we obtain
$\gamma_{11}=\gamma_{22}=\gamma_{33}$.  Thus, we have two coefficients
$\gamma_{00}=\Upsilon,\gamma_{ii}=-\Xi$.  Now, the Lagrangian

\[L_{0} = -(8\pi G)^{-1}\gamma_{\mu\nu}X^\mu_{,\alpha}X^\nu_{,\beta}
      	g^{\alpha\beta}\sqrt{-g}\]

fulfils this property.  For the difference $L-L_{0}$ we obtain

\[ {\delta \int(L-L_{0})\over\delta X^\mu} \equiv 0 \]

Thus, the remaining part is not only covariant in the weak sense, but
does not depend on the preferred coordinates $X^\mu$.  But this is
``strong'' covariance, the classical requirement for the Lagrangian of
general relativity.  Thus, we can identify the difference with the
classical Lagrangian of general relativity and obtain in the preferred
coordinates

\[L = -(8\pi G)^{-1}\gamma_{\mu\nu}g^{\mu\nu}\sqrt{-g}
    + L_{GR}(g_{\mu\nu}) + L_{matter}(g_{\mu\nu},\varphi^m)
\]

with the following modification of the Einstein equations

\[G^\mu_\nu  = 8\pi G (T_m)^\mu_\nu
   + (\Lambda +\gamma_{\kappa\lambda}g^{\kappa\lambda}) \delta^\mu_\nu
   - 2g^{\mu\kappa}\gamma_{\kappa\nu}\]

The last term is the full energy-momentum tensor, therefore, this
equations defines a decomposition of the energy-momentum tensor into
the energy-momentum tensor of matter and the energy-momentum
tensor of the gravitational field defined by

\[(T_g)^\mu_\nu = (8\pi G)^{-1}\left(\delta^\mu_\nu(\Lambda
              	+ \gamma_{\kappa\lambda}g^{\kappa\lambda})
	 	- G^\mu_\nu\right)\sqrt{-g}\]

\section{Quantization}

Most workers would agree that ``at the root of most of the conceptual
problems of quantum gravity'' is the idea that ``a theory of quantum
gravity must have something to say about the quantum nature of space
and time'' \cite{Butterfield}.  These problems, especially the problem
of time \cite{Isham}, simply disappear in a theory with fixed
Newtonian background.  Problems related with energy and momentum
conservation too -- the Hamiltonian is no longer a constraint.

The violation of Bell's inequality is independent evidence for a
preferred frame.  A preferred frame is required for compatibility with
the EPR criterion of reality \cite{EPR} and Bohmian mechanics
\cite{Bohm}.  Bell himself concludes \cite{Bell1}: ``the cheapest
resolution is something like going back to relativity as it was before
Einstein, when people like Lorentz and Poincare thought that there was
an aether --- a preferred frame of reference --- but that our
measuring instruments were distorted by motion in such a way that we
could no detect motion through the aether.''

Our theory is in ideal agreement with ``the present educated view on
the standard model, and of general relativity, ... that these are
leading terms in effective field theories'' \cite{Weinberg} -- an idea
introduced by Sakharov \cite{Sakharov}.  It seems natural to assume
that our medium has an atomic structure.  An interpretation of $\rho$
as the number of ``atoms'' per volume leads to an interesting
prediction for the cutoff:

\[ \rho(x) V_{cutoff} = 1. \]

It is non-covariant.  For the homogeneous universe, it seems to expand
together with the universe.  It differs from the usual expectation
that the cutoff is the Planck length $a_P\approx 10^{-33}cm$
(cf. \cite{Jegerlehner}, \cite{Volovik}).

\section{Comparison with other theories of gravity}

Because of the simplicity of the additional terms it is no wonder that
they have been already considered.  Two other theories have the same
Lagrangian for appropriate signs of the cosmological constants: the
``relativistic theory of gravity'' proposed by Logunov et
al. \cite{Logunov} and classical GR with some additional scalar
``dark matter'' fields.  Nonetheless, equations are not all.  There
are other physical important things which makes the theories different
as physical theories, like global restrictions, boundary conditions,
causality restrictions, quantization concepts which are closely
related with the underlying ``metaphysical'' assumptions.

\subsection{Comparison with RTG}

The ``relativistic theory of gravity'' (RTG) proposed by Logunov et
al. \cite{Logunov} has Minkowski background metric $\gamma_{\mu\nu}$.
The Lagrangian of RTG is

\[L = L_{Rosen} + L_{matter}(g_{\mu\nu},\psi^m)
    - {m_g^2}({1\over 2}\gamma_{\mu\nu}g^{\mu\nu}\sqrt{-g}
            - \sqrt{-g} - \sqrt{-\gamma})
\]

which de facto coincides with our theory for $\Lambda = - {m_g^2} <
0$, $\Xi = - \gamma^{11}{m_g^2} > 0$, $\Upsilon = \gamma^{00}{m_g^2} > 0$.

The metaphysical context of RTG is completely different.  It is a
special-relativistic theory, therefore incompatible with the EPR
criterion of reality and Bohmian mechanics.  Another difference is the
causality condition: In RTG, only solutions where the light cone of
$g_{ij}$ is inside the light cone of $\gamma_{ij}$ are allowed.  A
comparable but weaker condition exists in our theory too: $T(x)$
should be a time-like function, or, $\rho(X,T)>0$.  Note also that our
theory suggests a different way of quantization: the prediction for
the cutoff length $l_{cutoff}$ is not Lorentz-covariant.

\subsection{Comparison with GR plus dark matter}

The Lagrangian is also equivalent to GR with some dark matter -- four
scalar fields $X^{\mu}$.  In this theory they are no longer preferred
coordinates, but simply fields.  Such ``clock fields'' in GR have been
considered by Kuchar \cite{Kuchar}.  Usual energy conditions require
$\Xi>0, \Upsilon<0$.

This GR variant allows a lot of solutions where the fields
$X^{\mu}(x)$ cannot be used as global coordinates, especially
solutions with non-trivial topology.  They may also violate the
condition that $X^0(x)=T(x)$ is time-like.  Such solutions are
forbidden in our theory.  On the other hand, the infinite ``boundary
values'' of the ``fields'' $X^\mu(x)$ are unreasonable for matter
fields in GR.  Another difference is that in our theory the $X^\mu$
are fixed background coordinates and therefore should not be
quantized, while the ``fields'' $X^\mu(x)$ should be quantized.

\section{Predictions}

Using small enough values $\Xi, \Upsilon\to 0$ leads to GR equations.
Therefore it is not problematic to fit observation.  It is much more
problematic to find a way to distinguish our theory from GR by
observation.

\subsection{A dark matter candidate}

Let's consider the influence of the new terms on the expansion of the
universe.  In our theory a homogeneous universe is flat.  The the
usual ansatz $ds^2 = d\tau^2 - a^2(\tau)(dx^2+dy^2+dz^2)$ gives

\begin{eqnarray*}
3(\dot{a}/a)^2  &=&
   	- \Upsilon/a^6 + 3 \Xi/a^2 + \Lambda + \varepsilon\\
2(\ddot{a}/a) + (\dot{a}/a)^2 &=&
   	+ \Upsilon/a^6 +   \Xi/a^2 + \Lambda - p
\end{eqnarray*}

We see that $\Xi$ influences the expansion of the universe similar to
dark matter with $p=-{1\over 3}\varepsilon$.

\subsection{Big bounce instead of big bang singularity}

$\Upsilon$ becomes important only in the very early universe.  But for
$\Upsilon>0$, we obtain a qualitatively different picture. We obtain a
lower bound $a_0$ for $a(\tau)$ defined by

\[ \Upsilon/a_0^6 = 3 \Xi/a_0^2 + \Lambda + \varepsilon \]

The solution becomes symmetrical in time, with a big crash followed by
a big bang.  For example, if $\varepsilon = \Xi =0, \Upsilon>0,
\Lambda>0$ we have the solution

\[ a(\tau) = a_0 \cosh^{1/3}(\sqrt{3\Lambda} \tau)  \]

In time-symmetrical solutions of this type the horizon is, if not
infinite, at least big enough to solve the cosmological horizon
problem (cf. \cite{Primack}) without inflation.

\subsection{Frozen stars instead of black holes}

The choice $\Upsilon>0$ influences also another physically interesting
solution -- the gravitational collapse.  There are stable ``frozen
star'' solutions with radius slightly greater than their Schwarzschild
radius.  The collapse does not lead to horizon formation, but to a
bounce from the Schwarzschild radius.  Let's consider an example.  The
general stable spherically symmetric harmonic metric depends on one
step of freedom m(r) and has the form

\[ ds^2 = (1-{m \over r}{\partial m\over\partial r})
      	     	({r-m\over r+m}dt^2-{r+m\over r-m}dr^2)
       	- (r+m)^2 d\Omega^2 \]

Let's consider the ansatz $m(r)=(1-\Delta)r$. We obtain

\begin{eqnarray*}
ds^2 &=& \Delta^2dt^2 - (2-\Delta)^2(dr^2+r^2d\Omega^2) \\
0    &=& -\Upsilon \Delta^{-2} +3\Xi(2-\Delta)^{-2}+\Lambda+\varepsilon\\
0    &=& +\Upsilon \Delta^{-2} + \Xi(2-\Delta)^{-2}+\Lambda-p
\end{eqnarray*}

Now, for very small $\Delta$ even a very small $\Upsilon$ becomes
important, and we obtain a non-trivial stable solution for $p =
\varepsilon = \Upsilon g^{00}$.  Thus, the surface remains visible,
with time dilation $\sqrt{\varepsilon/\Upsilon}\sim M^{-1}$.

\end{document}